\RequirePackage{fix-cm}
\documentclass[smallcondensed]{svjour3}     
\smartqed  
\usepackage{graphicx}
\usepackage[cmex10]{amsmath}
\usepackage{enumerate}
\usepackage{subcaption}
\usepackage{nomencl}
\usepackage[ruled,vlined]{algorithm2e}

\usepackage{color}

\usepackage{soul}

\begin{document}

\title{Graph Embedding for Citation Recommendation
}


\author{Haofeng Jia         \and
        Erik Saule 
}


\institute{H. Jia \and E. Saule\at
           Dept. Computer Science\\
	   University of North Carolina at Charlotte\\
	   \email{\{hjia1,esaule\}@uncc.edu}
}


\maketitle

\begin{abstract}
As science advances, the academic community has published millions of research papers.
Researchers devote time and effort to search relevant manuscripts when writing a paper
or simply to keep up with current research. In this paper, we consider the problem of
citation recommendation on graph and propose a task-specific neighborhood construction strategy to
learn the distributed representations of papers. In addition, given the learned representations,
we investigate various schemes to rank the candidate papers for citation recommendation.
The experimental results show our proposed neighborhood construction strategy outperforms the widely-used
random walks based sampling strategy on all ranking schemes, and the model based ranking scheme
outperforms embedding based rankings for both neighborhood construction strategies.
We also demonstrated that graph embedding is a robust approach for citation recommendation when
hidden ratio changes, while the performance of classic methods drop significantly when the set of seed
papers is becoming small.

\end{abstract}

\section{Introduction}
\label{sec:intro}

Scientists around the world have published tens of millions of research papers, and the number of new papers
has been increasing with time. For example, according to DBLP~\cite{DBLP:journals/pvldb/Ley09}, computer scientists published 3 times more
papers in 2010 than in 2000. At the same time, literature search became an essential task performed daily by
thousands of researcher around the world. Finding relevant research works from the gigantic number of published
articles has become a nontrivial problem.\looseness=-1

Currently, many researchers rely on manual methods, such as keyword search via Google Scholar\footnote{https://scholar.google.com/} or Mendeley\footnote{https://www.mendeley.com/}, to discover new
research works. However, keyword based searches might not be satisfying for two reasons: firstly, the vocabulary gap
between the query and the relevant document might results in poor performance; secondly, a simple string of keywords
might not be enough to convey the information needs of researchers. There are many instances where such a keyword
query is either over broad, returning many articles that are loosely relevant to what the researcher actual need,
or too narrow, filtering many potentially relevant articles out or returning nothing at all~\cite{el2011beyond}.

To alleviate the above mentioned problems, many research works proposed citation recommendation algorithms which use a manuscript
instead of a set of keywords as query~\cite{strohman2007recommending,he2010context,he2011citation,lu2011recommending,huang2012recommending}. For example, context-aware citation recommendation is designed to recommend relevant papers for
placeholders in the query manuscript based on local contexts~\cite{he2010context,he2011citation}.
Manuscript based citation recommendation is great to help with the writing process. However, we are interested here in helping the research process which usually comes long before manuscripts are fleshed out.
Researchers have devoted efforts on citation recommendation based on a set of seed papers~\cite{mcnee2002recommending,torres2004enhancing,gori2006research,caragea2013can,Kucuktunc13-ASONAM}. Most approaches rely on the citation
graph to recommend relevant papers, such as collaborative filtering~\cite{mcnee2002recommending} and random walk framework~\cite{gori2006research}. The different approaches to recommending academic papers have been extensively surveyed by~\cite{Beel2016}.

We consider in this paper the problem of extending a set of known
references, which is helpful in recommender system and
academic search engine, such as theAdvisor~\cite{Kucuktunc13-JCDL}.
We explore the node embedding on graph for this citation recommendation task.
This work is credited to recent development on language model and graph embedding. In the context of word embedding,
the notion of neighborhood can be defined using a sliding window over consecutive words. While in the
context of graph embedding, nodes are not linearly structured, so before moving to the embedding model phase,
we need a strategy to sample nodes sequences like the sentences in natural language then feed them to the model.
This sampling process is called context/neighborhood construction from graph.

It turns out that the way to define neighborhood is critical and can significantly affect the performance.
Streams of short random walks is becoming a popular way to build the neighborhood~\cite{perozzi2014deepwalk,grover2016node2vec,dong2017metapath2vec,jiang2018cross,ganguly2017paper2vec}.
In this paper, we propose
a strategy using co-citation based sampling. The experimental results show the proposed
sampling strategy outperforms the random walks based sampling strategy on citation recommendation
task.

Given the learned representations,
we investigate various schemes to rank the candidate papers for citation recommendation.
The results show the model based ranking scheme outperforms embedding based rankings for both sampling strategies.

Then we show that graph embedding is a robust approach for citation recommendation when hidden ratio changes
compared with classic methods and when the size of seed paper set is small, co-citation sampling based embedding
is a better choice.

To summarize, our major contributions in this paper are as follows:
\begin{enumerate}
\item We evaluate random-walk neighborhood construction strategy on citation recommendation task.

\item We propose a task-specific neighborhood construction strategy for citation recommendation.

\item We investigate two ranking schemes for citation recommendation: embedding based and model based.

\item We show that the proposed approaches outperforms classic approaches for those queries with a few seed papers.

\end{enumerate}

The paper is organized as follows: we introduce the problem statement and related work in Sec.~\ref{sec:relwork}. In
Sec.~\ref{sec:learning}, we introduce the neural learning framework for citation recommendation.
Then we explore sampling strategies and ranking schemes in Sec.~\ref{sec:sampling} and
Sec.~\ref{sec:ranking} respectively. Sec.~\ref{sec:embexp} shows the experimental results. Finally,
Sec.~\ref{sec:discussion} discuss the usefulness of graph embedding on citation recommendation task and
Sec.~\ref{sec:ccl} concludes this paper and discuss our future work.

\section{Problem Definition and Related Work}
\label{sec:relwork}
In this section, we first formalize the problem of citation recommendation based on a set of
seed papers and then introduce the related literatures.
\subsection{Problem Definition}
Let $G=(V,E)$ be the citation graph, with $n$ papers $V=\{v_1,\ldots,v_n\}$. In $G$, each
edge $e\in E$ represents a citation relationship between two papers.
We use $Ref(v)$ and $Cit(v)$ to denote the reference set of and citation set to $v$,
respectively. And $Adj(v)$ is used to denote the union of $Ref(v)$ and $Cit(v)$.

In this paper, we focus on citation recommendation problem assuming that researchers already
know a set of relevant papers.  Therefore, the task can be formalized as:

\emph{Citation Recommendation.} Given a set of seed papers $S$,
return a list of papers ranked by relevance to the ones in $S$.


\subsection{Citation Recommendation}

Academic citation recommendation methods can be classified based on
the input that the recommendation system takes. There have been an
interest in methods that recommend academic papers based on an
existing manuscript, typically taking an existing manuscript as an
input to annotate~\cite{he2010context,he2011citation}. The usage of
metadata, such as authors or keywords, have been considered to build
discriminative models~\cite{yu2012citation,liu2014full,liu2014meta,ren2014cluscite}. The
problem of the technical vocabulary changing depending on communities
or application area has spawned works to bridge the gap between two
heterogeneous
languages~\cite{lu2011recommending,huang2012recommending}. The effects of modeling a
researcher's past works and exploiting potential citation papers in recommending papers are also examined~\cite{sugiyama2010scholarly,sugiyama2013exploiting}.
Most notably this stream of research as generated a tool called
RefSeer\footnote{http://refseer.ist.psu.edu/}~\cite{huang2014refseer}
which suggests recommendations on a paragraph or manuscript using
topic-based global recommendations and citation-context-based local
recommendations.

While manuscript based recommendation is a worthwhile problem, we focus in this paper on
\emph{seed papers based citation recommendation.}
Given a "basket" of citations, McNee et al.~\cite{mcnee2002recommending} explore the use of
Collaborative Filtering (CF) to recommend papers that would be suitable additional references
for a target research paper. They create a ratings matrix where citing papers correspond to
users and citations correspond to items. The experiments show CF could generate high quality
recommendations.
As a follow-up, Torres et al.~\cite{torres2004enhancing} describe and test different techniques
for combining Collaborative Filtering and Content-Based Filtering.
A user study shows many of CF-CBF hybrid recommender algorithms can generate research paper
recommendations that users were happy to receive. However, offline experiments show those
hybrid algorithms did not perform well. In their opinion, the sequential nature of
these hybrid algorithms: the second module is only able to make recommendations seeded by the
results of the first module. To address this problem, Ekstrand et al.~\cite{ekstrand2010automatically}
propose to fuse the two steps by running a CF and a CBF
recommender in parallel and blending the resulting ranked lists. The first items on the combined
recommendation list are those items which appeared on both lists, ordered by the sum of
their ranks. Surprisingly, Collaborative Filtering outperforms
all hybrid algorithms in their experiments.

Gori et al.~\cite{gori2006research} devised PaperRank, a random walk based method
to recommend papers according to a small set of user selected relevant articles.
K\"u\c{c}\"uktun\c{c} et al. designed a personalized paper recommendation service, called theAdvisor\footnote{http://theadvisor.osu.edu/}~\cite{Kucuktunc13-ASONAM,Kucuktunc13-JCDL},
which allows a user to specify her search toward recent developments or traditional papers using a
direction-aware random walk with restart algorithm~\cite{Kucuktunc12-DBRank}.  The recommended papers
returned by theAdvisor are diversified by parameterized relaxed local maxima~\cite{Kucuktunc13-WWW}.
K\"u\c{c}\"uktun\c{c} et al. proposed sparse matrix ordering and partitioning techniques to accelerate citation such recommendation
algorithms~\cite{Kucuktunc12-ASONAM}.

Caragea et al.~\cite{caragea2013can} addressed the problem of citation recommendation
using singular value decomposition on the adjacency matrix associated with the citation graph
to construct a latent semantic space:
a lower-dimensional space where correlated papers can be  easily identified. Their experiments
on Citeseer show this approach achieves significant success compared with Collaborative
Filtering methods.
Wang et al.~\cite{Wang2011} proposes to include textual information to build an topic model of the papers and adds an additional latent variable to distinguish between the focus of a paper and the context of the paper.\looseness=-1

A typical related paper search scenario is that a user starts with a seed of one or more papers,
by reading the available text and searching related cited references. Sofia is a system
that automates this recursive process \cite{golshan2012sofia}.

The approach proposed by El-Arini and Guestrin~\cite{el2011beyond} returns a set of relevant articles by optimizing
a function based on a fine-grained notion of influence between documents; and also claim that, for paper recommendation, defining a query as a small set of known-to-be-relevant
papers is better than a string of keywords.

The graph embedding methods that we will present use sampling method
inspired by the classic seed papers based citation recommendation
PaperRank~\cite{gori2006research} and Collaborative
Filtering~\cite{mcnee2002recommending}. And we will also evaluate the
recommendation results our graph embedding methods obtain against
these two baselines.

\subsection{Graph Embedding}

This work is also credited to recent development on language model~\cite{bengio2003neural,mikolov2013efficient,mikolov2013distributed,le2014distributed}
and graph embedding~\cite{perozzi2014deepwalk,tang2015line,grover2016node2vec,liu2016graph,goyal2017graph,wang2017community,huang2017label,dong2017metapath2vec,cai2018comprehensive} which represent words or vertices as vectors in a low dimensional space.
On the one hand, in the language model, each word is represented by
a vector which is concatenated or averaged with other word
vectors in a context, and the resulting vector is used to predict
other words in the context. For example, the neural
network language model proposed by Bengio et al.~\cite{bengio2003neural}
uses the concatenation of several previous word vectors to
form the input of a neural network, and tries to predict the
next word. The outcome is that after the model is trained,
the word vectors are mapped into a vector space such that
semantically similar words have similar vector representations.

On the other hand, the problem of graph embedding is related to two
traditional research problems, i.e., graph analysis and
representation learning. Particularly, graph embedding
aims to represent a graph as low dimensional vectors while
the graph structures are preserved. Graph
analysis aims to mine useful information from graph data.
And representation learning obtains data
representations that make it easier to extract useful information
when building classifiers or other predictors. Graph
embedding lies in the overlap of the two problems and
focuses on learning the low-dimensional representations.
Recently, deep learning (unsupervised feature learning) techniques,
which have proven successful in natural language processing, has been
introduced for graph analysis. For example, DeepWalk~\cite{perozzi2014deepwalk}
learns social representations of a graph's vertices, by modeling
a stream of short random walks. Social representations
are latent features of the vertices that capture neighborhood
similarity and community membership. These latent
representations encode social relations in a continuous
vector space with a relatively small number of dimensions.

Since then, Random walk sampling has become the most popular
neighborhood construction strategy for graph embedding. Node2vec~\cite{grover2016node2vec}
extends the DeepWalk by leveraging breadth first sampling and depth first sampling and
Metapath2vec~\cite{dong2017metapath2vec} utilizes the random walk sampling on heterogeneous graphs.
Efforts have been made to bring the idea the citation recommendation related field. Jiang et al.~\cite{jiang2018cross}
explore cross language citation recommendation by guiding the random walk streams. Paper2vec~\cite{ganguly2017paper2vec}
extends the edges in citation graph based on textual similarities, then adopts the random walk sampling to learn the representations
of papers in the graph. In this paper, we propose a novel strategy for neighborhood construction for citation recommendation on graph.

Recent advancements in representation learning methods have
proven to be effective in modeling distributed representations
in different modalities like images, languages, speech,
graphs etc. The distributed representations obtained using
such techniques in turn can be used to calculate similarities.

The techniques presented in this paper adapt the classic language
model and graph embedding techniques discussed here and extend them
to solve the problem of seed paper based citation recommendation by
embedding an academic citation graph.

\section{Learning Framework}
\label{sec:learning}
\begin{figure}[tbp]
  \begin{center}
    \includegraphics[width=0.8\linewidth]{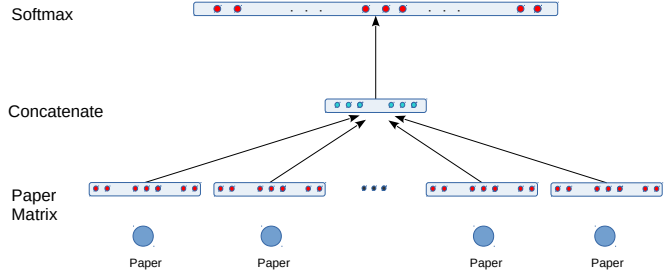}
  \end{center}
  \vspace{-3.9em}
  \caption{Learning Framework}
  \label{fig:learning}
\end{figure}
In this section, we introduce the framework to learn representation vectors
for papers from the citation graph. Since our target is recommending citations
based on a set of know papers, we extend the continuous bag-of-word architecture
for our problem.

Formally, we aim to learn a mapping function that maps papers to distributed
representations. To this end, we maximize the log probability of observing a paper
conditioned on a set of neighborhood papers:

$$max\sum_{p \in V}{\log Pr(p|Neib(p))} $$

$Neib(p)$ consists of a set of neighborhood papers of paper $p$ and they are
generated through a neighborhood sampling strategy.

Figure~\ref{fig:learning} shows the learning architecture. The input layer consists of
a set of papers, which are the neighborhood papers of the target paper. The parameter matrix
between the input layer and hidden layer is essentially the representation matrix, through
which we obtain the distributed representations for input papers. The hidden layer is typically
computed as the aggregated representation for input papers, then through the parameter matrix
between hidden layer and output layer, we obtain the predicted probabilities for every paper
conditioned on the input papers, and the softmax function is used here:

$$Pr(p|Neib(p)) = \frac{e^{y_{p}}}{\sum_{i}{e^{y_{i}}}}$$

Each of $y_{i}$ is the un-normalized probability for a candidate
paper $i$.

We train this model on a bunch of paper sequences generated from the citation graph, and stochastic
gradient descent is employed to optimize the parameter matrices. Defining the neighborhood of a paper
in the citation graph is critical, therefore we investigate the neighborhood construction strategies in
the following part.
\section{Neighborhood Construction Strategies}
\label{sec:sampling}
In the context of word embedding, the notion of neighborhood can be defined using
a sliding window over consecutive words, because of the linear nature of nature
language text. However, papers in a citation graph are not linearly structured.
In this section, we propose different strategies to define the notion of a neighborhood of
a source paper.
\subsection{Random Walk Stream}
Generating streams of short random walks is a popular way to linearize the node
relationships in graph~\cite{perozzi2014deepwalk,grover2016node2vec,dong2017metapath2vec,jiang2018cross,ganguly2017paper2vec}. Random walks have been used
as a similarity measure for a variety of problems in content recommendation
and community detection. They are also the foundation of a class of
output sensitive algorithms which use them to compute local community structure
information in time sublinear to the size of the input graph. Recent work
has shown the ability of random walk to learn social representations of vertices
in social graphs~\cite{perozzi2014deepwalk}.

Formally, we denote a random walk rooted at node $v_i$ as $W_{v_i}$.
It is a stochastic process with random variables $W_{v_i}^{1}$,$W_{v_i}^{2}$,...,$W_{v_i}^{t}$
such that $W_{v_i}^{k+1}$ is a node chosen uniformly at random from the immediate adjacent neighbors
of node $v_{k}$ in the graph.  We start the random walk generation with the fixed walk length at
each node respectively. And in order to obtain robust embedding, we repeat the above process for a number of times.

As we shown in Algorithm~\ref{alg1}, the outer loop iterates $n$ times, each iteration is making a pass
over the graph and sample one walk stream per node during this pass. For the inner loop,
all nodes of the graph are traversed and a random walk stream with length $t$ is sampled at each node.
The sampling process starting with node $v$ is described in Algorithm~\ref{alg2}.

\begin{algorithm}[H]
\label{alg1}
\SetAlgoLined

 \For{$iter =1$ to $n$}{
  Shuffle($V$)\;
  \ForEach{$v \in V$}{
    $\mathcal{RWS}$ =RandomWalkSampling($G,v,t$)\;
    Append $\mathcal{RWS}$ to $walks$\;
   }
 }
 return $walks$
 \caption{Random Walks Generation Process}
\end{algorithm}

\begin{algorithm}[H]
\label{alg2}
\SetAlgoLined
 Initialize $walk$ to $[v]$\;
 \For{$iter =1$ to $t$}{
   $walk[iter]$ = PickNeighborOf($walk[iter-1]$)
 }
 return $walk$
 \caption{RandomWalkSampling($G,v,t$)}
\end{algorithm}

The object of the model is to estimate the likelihood of observing vertex $v_i$ given all the
previous vertices visited so far in the random walk.
$$Pr(v_i|(v_1,v_2,...,v_{i-1}))$$

A stream of short random walks can capture the local structure
information and this model is easy to parallelize. Several random
walkers in different threads, processes, or machines can simultaneously
explore different parts of the same graph.

The above random walk sampling strategy uniformly picks a neighbor at random. A variant is
to try to interpolate between breadth first search and depth first search~\cite{grover2016node2vec}.
The breadth first and depth first sampling represent two extreme scenarios
in terms of the search space. In Breadth first sampling, the neighborhood
is restricted to nodes which are immediate neighbors of the source. On the contrary,
in Depth first sampling, the neighborhood consists of nodes sequentially sampled
at increasing distances from the source node. The intuition behind this two sampling
schemes is that they can capture two kinds of node similarities: homophily and structural
equivalence. Under the homophily hypothesis~\cite{fortunato2010community,yang2014overlapping} nodes
that are highly interconnected and belong to similar network clusters
or communities should be embedded closely together.
While under the structural equivalence hypothesis~\cite{henderson2012rolx}
nodes that have similar structural roles in networks should be embedded
closely together.

In order to allow us to account for the graph structure and guide our
search procedure to explore different types of network neighborhoods and
interpolate between breadth first sampling and depth first sampling. A
search bias $\alpha$ is introduced.

A second order random walk is guided with two parameters $p$ and $q$.
Let use assume a random walk that just traversed
from node $t$ to node $v$ and now resides at node $v$. The next step of
the walk is decided on the transition probabilities:
$$\pi_{vx}\propto\alpha_{pq}(t,x)\times weig_{vx}$$.
The transition probability from $v$ to $x$ is proportional to edge weight $weig_{vx}$ and
the search bias $\alpha_{pq}(t,x)$, which is defined as follows:
\[
  \alpha_{pq}(t,x) =
  \begin{cases}
     \frac{1}{p}  & \text{if } d_{tx}=0 \\
     1 & \text{if } d_{tx}=1 \\
     \frac{1}{q}  & \text{if } d_{tx}=2 \\
  \end{cases}
  \]
where $d_{tx}$ mean the distance between $t$ and $x$ in the graph.

In our task, citation graph is unweighted ($weig_{vx} = 1$), so we only care
about the search bias $\alpha$. Parameter $p$ controls the likelihood of immediately
revisiting a node in the walk. Setting it to a high value
ensures that we are less likely to sample an already visited
node in the following two steps. This strategy encourages moderate
exploration and avoids 2-hop redundancy in sampling. On the
other hand, if $p$ is low, it would lead the walk to
backtrack a step and this would keep the walk local
close to the starting node.

Parameter $q$ allows the search to differentiate
between inward and outward nodes. If $q > 1$, the random walk is biased
towards nodes close to node t.
Such walks obtain a local view of the underlying graph with respect
to the start node in the walk and approximate BFS behavior in the
sense that our samples comprise of nodes within a small locality.
In contrast, if $q < 1$, the walk is more inclined to visit nodes
which are further away from the node t. In particular, DeepWalk~\cite{perozzi2014deepwalk}
is a special case of Node2vec~\cite{grover2016node2vec}, where parameter $p = 1$ and
$q = 1$.

\subsection{Co-Citation Context}

Random walk based sampling strategies can encode homophily and structural similarities
in some extent. However, for a specific task where we want to sample sequences that capture
a certain property, those similarities seem too general to achieve a good performance.
For citation recommendation, we care more about the co-cited relationship. Under this
circumstance, random walk based samplings tend to bring noise, especially for those nodes
with a large number of citations.

\begin{figure}[tbp]
  \begin{center}
    \includegraphics[width=0.8\linewidth]{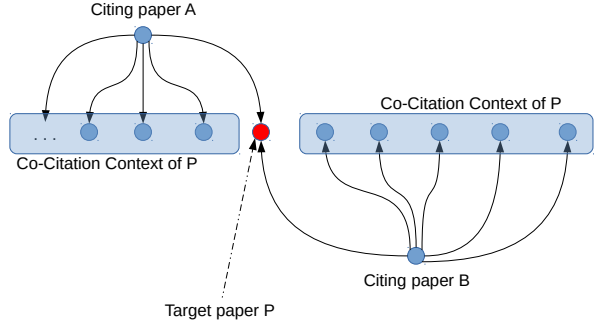}
  \end{center}
    \vspace{-3.9em}
  \caption{Context Construction}
  \label{fig:context}
\end{figure}

Classic word2vec~\cite{mikolov2013distributed} constructs the context of a word as the words which co-occur
with the target word within a sliding window. In the citation recommendation task,
we consider the context of a paper as other papers which co-occur
in one of its citing paper's reference list. For example, in Figure~\ref{fig:context}, the
context of target paper $P$ is highlighted by blue rectangles.

We consider a sampling strategy that emphasizes similarity between those co-cited papers.
Taking co-citation papers as the context of target paper seems
intuitive and reasonable for citation recommendation task.

As we shown in Algorithm~\ref{alg3}, The outer loop iterates $n$ times, each iteration
is making a pass over the graph and shuffle the reference list for each node during this
pass. For the inner loop, all nodes of the graph are traversed and we append the shuffled
reference list to $walks$.

\begin{algorithm}[H]
\label{alg3}
\SetAlgoLined

 \For{$iter =1$ to $n$}{
  Shuffle($V$)\;
  \ForEach{$v \in V$}{
    $\mathcal{RL}$=Shuffle(ReferenceList($v$))\;
    Append $\mathcal{RL}$ to $walks$\;

   }
   }

 return $walks$
 \caption{Co-Citation Sampling Process}
\end{algorithm}

\section{Ranking Strategies}
\label{sec:ranking}
After the embedding model is trained, we need to find a way to rank all the candidate papers.
The first idea is based on the learned distributed representation of papers, which we can obtain
from the weighting matrix between the input layer and the hidden layer after the training is finished.
An alternative strategy is using the trained model to predict the probabilities of candidate papers appearing
in the context of seed papers, where both the weighting matrix between the input layer and the hidden layer and
the weighting matrix between the hidden layer and the output layer are involved.
\subsection{Embedding Based Ranking}
Given the learned distributed representation of papers, we design three different
approaches to score the candidate papers based on a set of seed papers.

In the first one, we calculate the cosine similarities between the candidate paper $d$ and all seed
papers in $S$, then the average value, which is denoted as $simAvg$, is used to rank all candidate papers.
$$SimAvg_{d} = \frac{\sum_{s \in S} Cos(E_{s},E_{d})}{|S|} $$
where $E_{d}$ means the embedded vector of node $d$ in the citation graph.

We also consider the fact that seed papers might not contribute equally to finding hidden papers.
So we derive weights of seed papers in inverse proportion to their degrees.
$$SimWgd_{d} = \frac{\sum_{s \in S} \frac{1}{\delta_{s}} Cos(E_{s},E_{d})}{|S|} $$
where $\delta_{s}$ denotes the degree of seed paper $s$ in the citation graph.

Another metric firstly computes the average of seed papers as a reference paper, then the cosine
similarity between the reference paper and candidate paper $d$ is taken as $SimRef$.
$$SimRef_{d} = Cos \left ( \frac{\sum_{s \in S}E_{s}}{|S|},E_{d} \right ) $$

In general, embedding based ranking calculates the second order proximity of nodes in graph.
For many applications, the hidden layer and output layer are discarded once the training is finished,
since the aim of embedding is only to obtain the distributed representation.
\subsection{Model Based Ranking}
In order to rank candidate papers in a reasonable way, we also consider to use the trained model to predict
the probabilities of candidate papers appearing in the context of seed papers. This model based ranking strategy
measures the first order proximity of nodes in graph.

In specific, we first aggregate the distributed representations of seed papers, then multiply by the weighting matrix
between hidden layer and output layer, and use softmax to normalize the probabilities in the output vectors.
Those probabilities are then used to rank corresponding candidate papers. We denote this model based ranking as $CitMod$.

Essentially, the embedding based ranking strategy scores the candidate papers based on their similarities to seed papers,
while the model based ranking strategy is based on their relevance to seed papers. In next section, we
show the comparison experiments for various sampling and ranking strategies.
\section{Experiment}
\label{sec:embexp}
\subsection{Data Preparation}
To obtain a clean and comprehensive academic data set, we match
Microsoft Academic Graph\footnote{https://www.microsoft.com/en-us/research/project/microsoft-academic-graph/}~\cite{sinha2015overview},
CiteSeerX\footnote{http://citeseerx.ist.psu.edu/} and DBLP\footnote{http://dblp.uni-trier.de/xml/}~\cite{DBLP:journals/pvldb/Ley09}
datasets for their complementary advantages and derive a corpus of
Computer Science papers. Finally, we obtain a citation graph with 2,035,246 papers and 12,439,090 citations.

\subsection{Experimental Setup}
In order to simulate the typical use case where a researcher is writing
a paper and tries to find some more references, we design the random-hide
experiment.  Instead of removing all the irrelevant papers from the citation graph (to simulate the time
when the query paper was being written) when a query comes in, we train distributed representations
of nodes every year between 2004 to 2009. For example, we remove all papers published after 2006
from the citation graph to get Graph-in-2006 and generate $walks$ from Graph-in-2006 and train
embedding for nodes in that graph. In this way, we can obtain different graph embeddings from 2004 to 2009.
Query papers with 20 to 200 references and published between 2005 to 2010 are randomly (uniformly) selected from the dataset.
For each query paper, the embedding before the publishing year is
used for the task. Then, we randomly hide 10\% of the references as hidden set.
This set of hidden paper is used as ground truth to recommend. The remaining papers are used as
the set of seed papers.

Finally, to evaluate the effectiveness of recommendation algorithm, we use \emph{recall@$k$},
the ratio of hidden papers appearing in top $k$ of the recommended list. Performance on
average recall for 2,500 independent randomly selected queries is used for evaluation.

Random walk sampling strategy involves a number of parameters. In the following experiments, default
number of walks $n$ and walk length $t$ are set to 10 and 80 respectively, the parameters $p$ and $q$
are both set to $1$\footnote{Though we examined different values, the differences on performance are marginal}.
During training process, the dimension of embedded space is set to 128 and
the window size of neighborhood is set to 10 by default. For co-citation sampling, number of walks,
dimension of embedded space and window size are set to the same value as Random walk sampling strategy.

\subsection{Comparison of the Different Embedding Based Methods}

\begin{figure}[tbp]
\centering
\begin{subfigure}{0.49\linewidth}
  \centering
  \includegraphics[width=\textwidth]{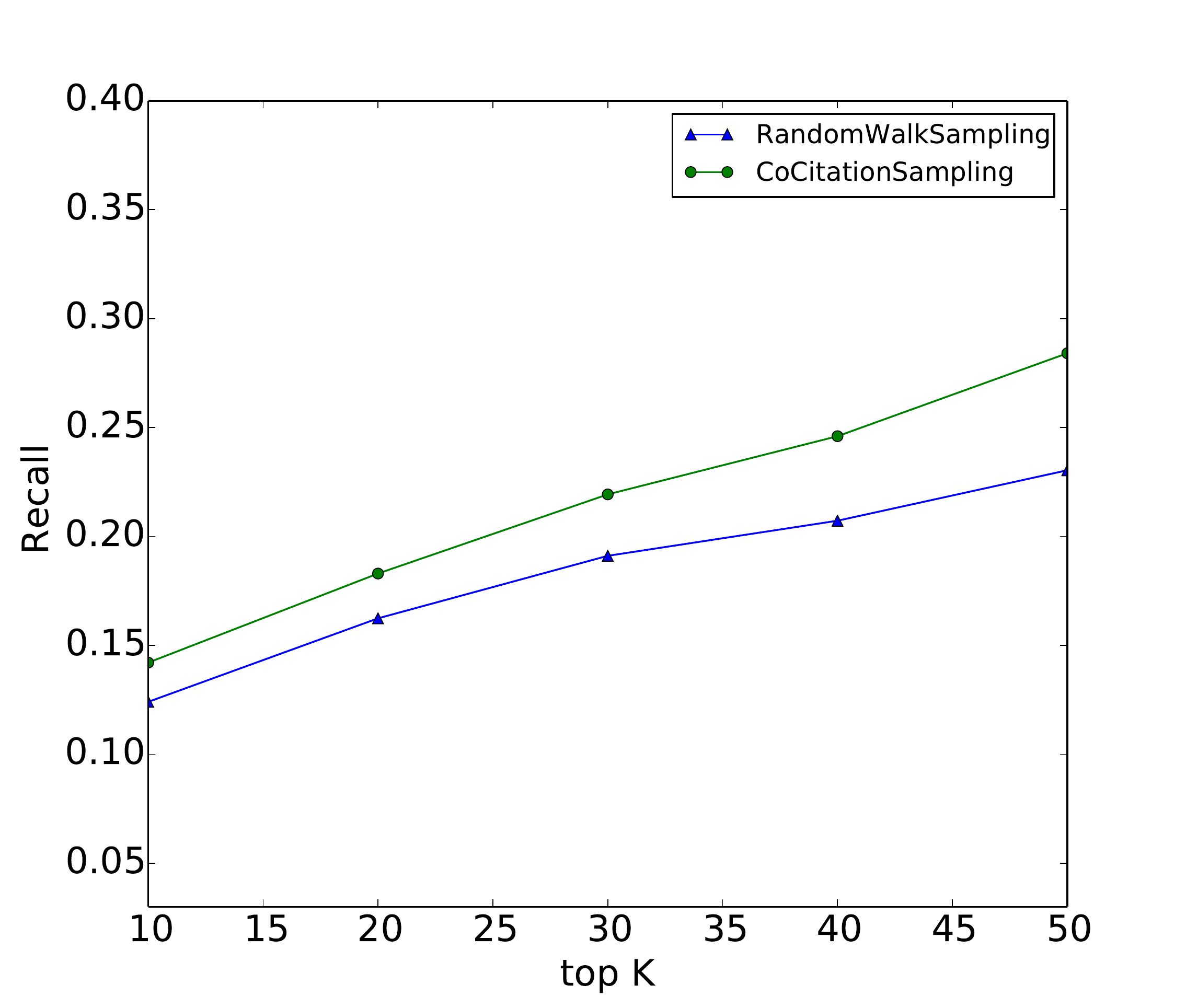}
  \caption{SimAvg}
\end{subfigure}
\begin{subfigure}{0.49\linewidth}
  \centering
  \includegraphics[width=\textwidth]{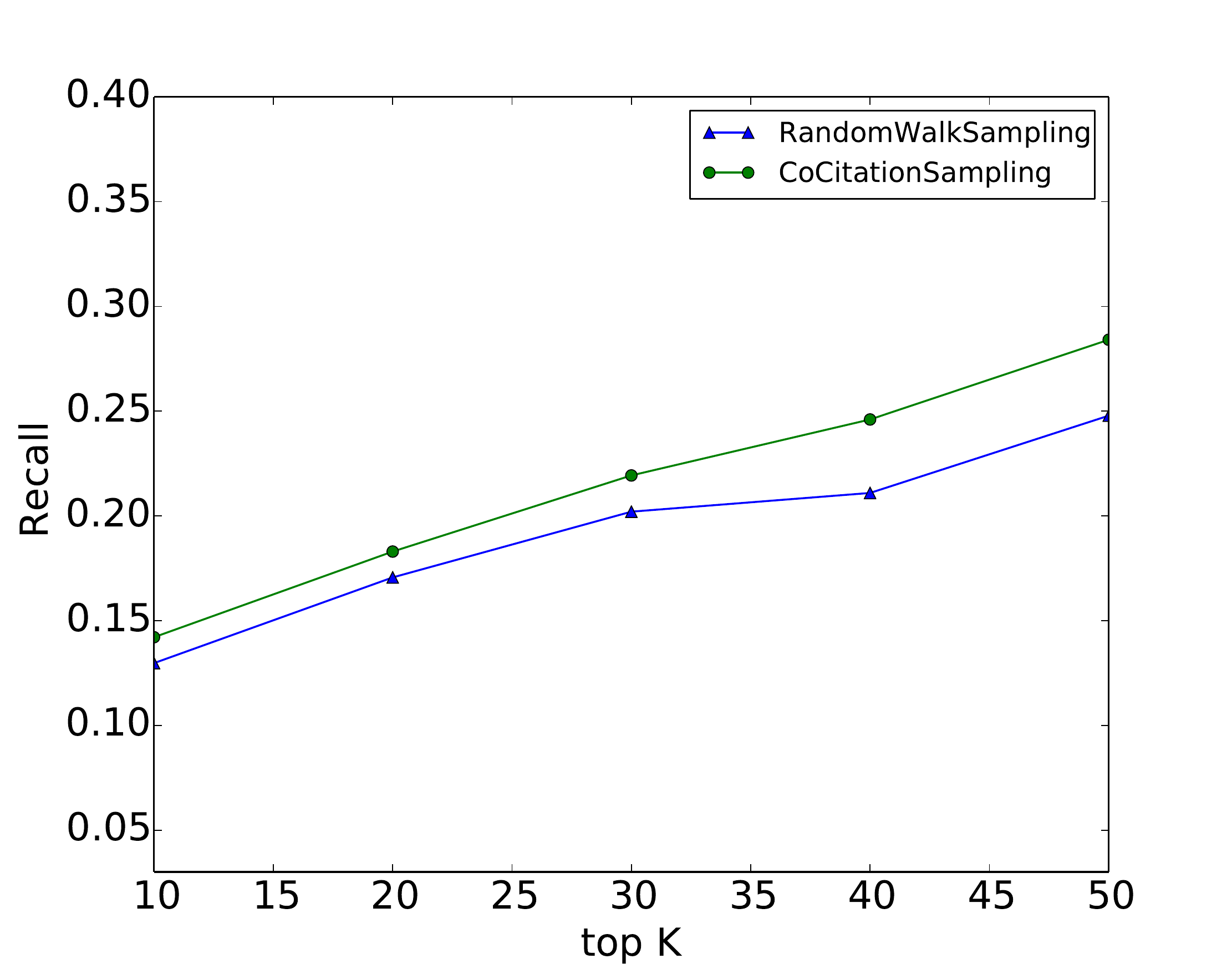}
  \caption{SimWgd}
\end{subfigure}
\begin{subfigure}{0.49\linewidth}
  \centering
  \includegraphics[width=\textwidth]{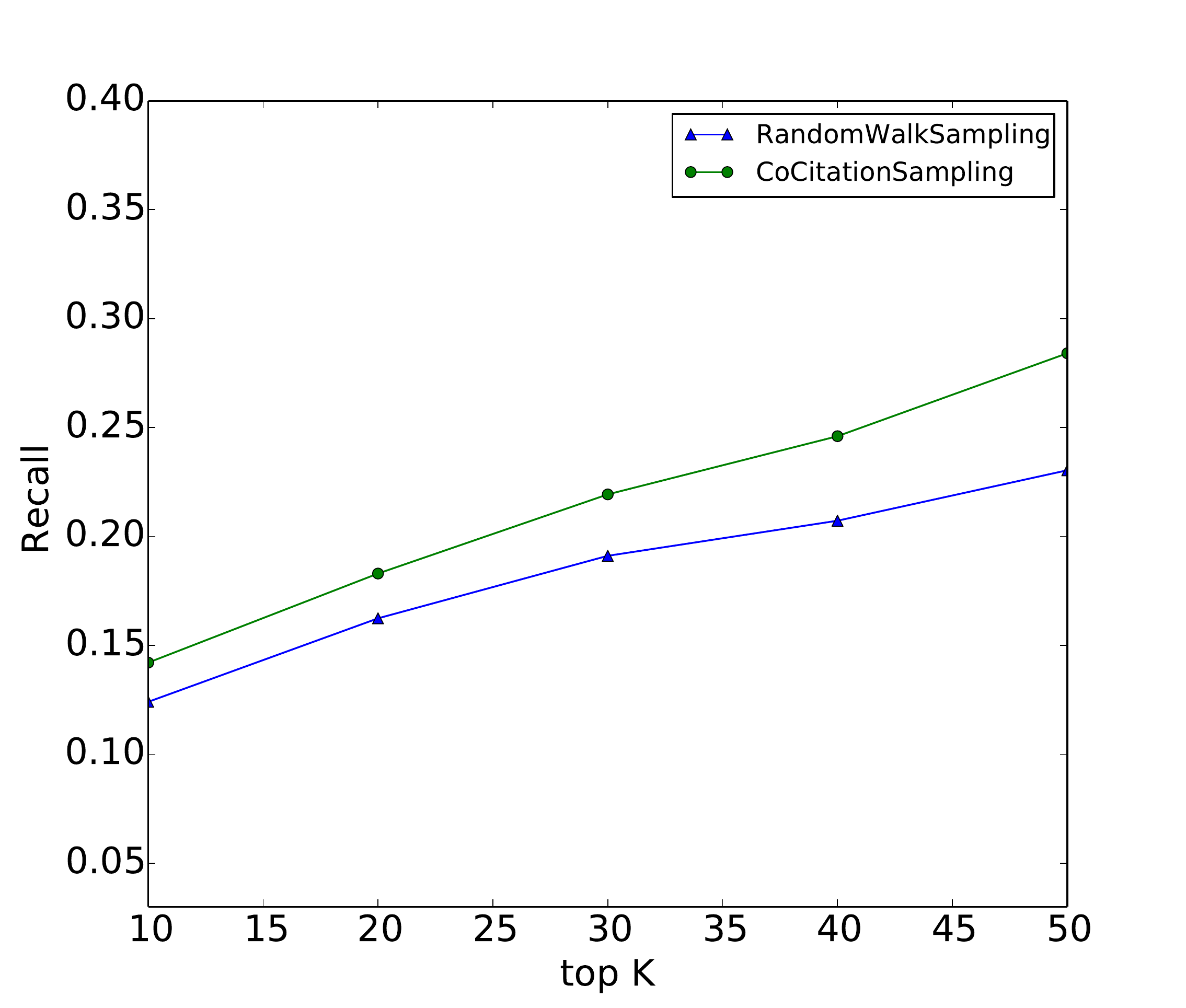}
  \caption{SimRef}
\end{subfigure}
\begin{subfigure}{0.49\linewidth}
  \centering
  \includegraphics[width=\textwidth]{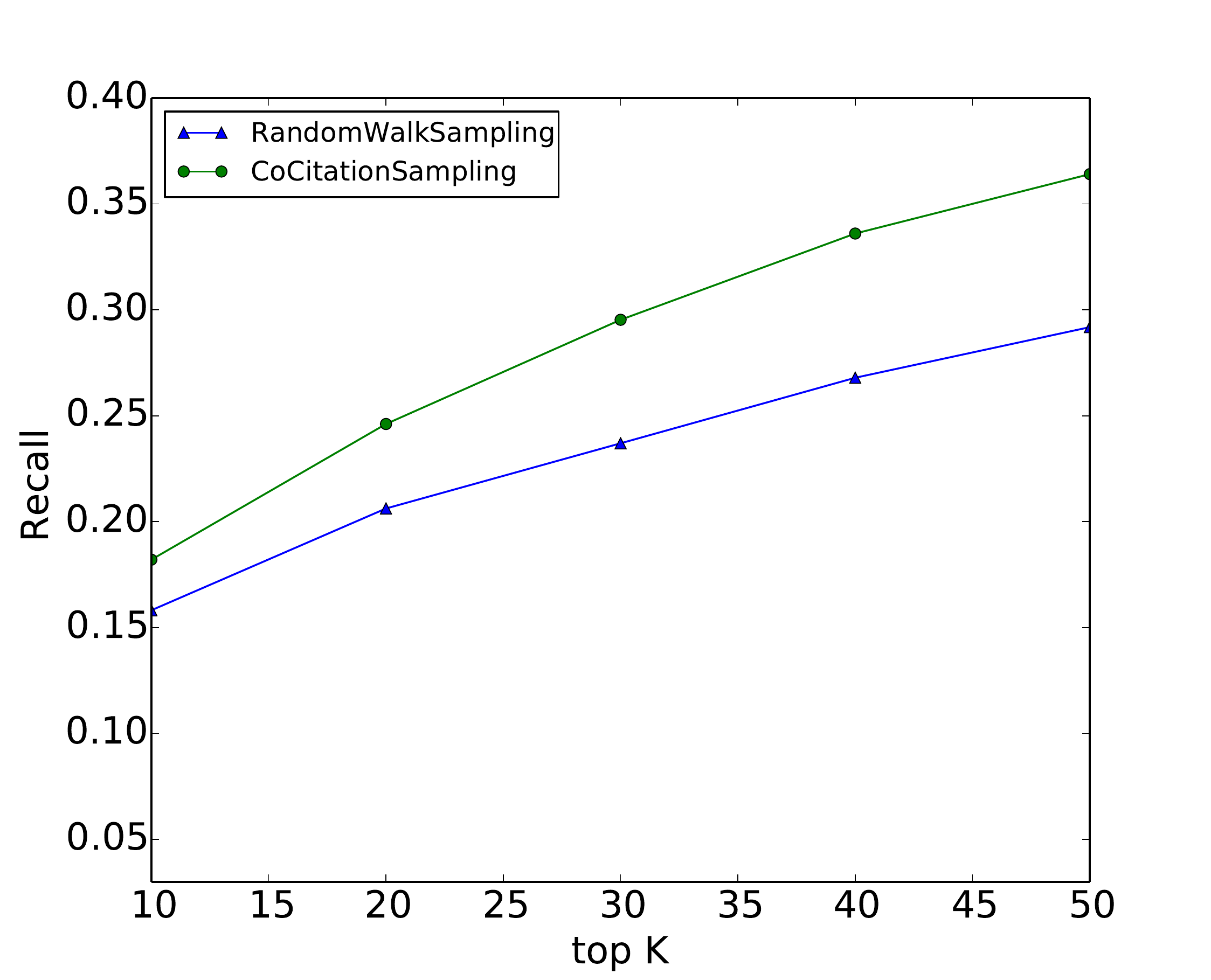}
  \caption{CitMod}
\end{subfigure}

\caption{Performance Comparison for Different Neighborhood Construction Strategies}
\label{fig:perfofsampling}
\end{figure}

Figure~\ref{fig:perfofsampling} shows the performance of different context sampling strategies on
citation recommendation. Generally speaking, co-citation based sampling is achieving better recall than
random walk based sampling on all ranking schemes, and the model based ranking $CitMod$ outperforms all three
embedding based rankings. Surprisingly the $SimAvg$ performs similarly as $SimRef$,
which firstly compute a reference paper by averaging seed papers.

In particular, for $SimAvg$ and $SimRef$ co-citation sampling is 14.52\% higher than random walk sampling on $recall@10$ and 23.47\%
higher on $recall@50$; for $SimWgd$ co-citation sampling is 10.48\% higher than random walk sampling on $recall@10$ and 14.98\%
higher on $recall@50$. In other words, the weighted approach $SimWgd$ brings higher impact on random walk sampling compared with co-citation
sampling. The model based ranking $CitMod$ outperforms embedding based rankings by at least 28.17\% and 22.48\% for co-citation sampling and
random walk sampling respectively.

\subsection{Usefulness in Real Systems and Comparisons  to State of the Art}
\label{sec:discussion}
In the experimental setup, for each query paper we randomly hide 10\% of its references as hidden set.
This set of hidden paper is used as ground truth to recommend. The remaining papers are used as
the set of seed papers. This experiment is designed to simulate the scenario that when a researcher
want to explore more papers based on a set of seed papers. There are many cases that seed papers are
small. Here we define the ratio of hidden set out of the reference list as \emph{hidden ratio}. And we
examine how different hidden ratios affect the performance.

We compare the co-citation sampling (CCS) based embedding with classic
citation recommendation methods PaperRank\cite{gori2006research} and CF\cite{mcnee2002recommending} on different hidden ratios.
As we can see in Table~\ref{tab:low_hidden}, where hidden ratios are low, both PaperRank and CF are significant better than CCS.
However, the performance decreases as the hidden ratio increases for PR and CF, while CCS seems to be robuster. Table~\ref{tab:high_hidden}
shows their performance for high hidden ratios, and the proposed methods outperform classic methods, which means the proposed
methods are suitable for the cases that the size of input set is small.

In particular, when the hidden ratio increases from 10\% to 95\%, the the performance of co-citation sampling based
embedding decreases by 12.92\%, while the performance of PR and CF decreases by 46.95\% and 38.22\% respectively.
CCS reaches a better performance than both PR and CF when the hidden ratio is large.

This experiment demonstrates CCS is a robust approach for citation recommendation when hidden ratio changes, while
the performance of PR and CF drops a lot when hidden ratio is becoming large. In general, when the hidden ratio is
small, classic methods are better, but when the size of seed paper set is small, co-citation sampling based embedding
is a better choice.
\begin{table}[tbp]
\caption{Recall for Low Hidden Ratios}
\centering
\begin{tabular}{c|rrr}
 \hline
  Hidden Ratio & 10\% & 20\% \\\hline
  \rmfamily PaperRank & \rmfamily \textbf{0.234413}  & \rmfamily \textbf{0.224196} \\
  \rmfamily CF & \rmfamily 0.191736  & \rmfamily 0.186165 \\
  \rmfamily CoCitationSampling+SimAvg & \rmfamily 0.147815  & \rmfamily 0.145573 \\
  \rmfamily CoCitationSampling+CitMod & \rmfamily 0.182156  & \rmfamily 0.179513 \\
  \hline
\end{tabular}
\label{tab:low_hidden}
\end{table}
\begin{table}[tbp]
\caption{Recall for High Hidden Ratios}
\centering
\begin{tabular}{c|rrr}
 \hline
  Hidden Ratio & 90\% & 95\% \\\hline
  \rmfamily PaperRank & \rmfamily 0.134510  & \rmfamily 0.122956 \\
  \rmfamily CF & \rmfamily 0.119756  & \rmfamily 0.112389 \\
  \rmfamily CoCitationSampling+SimAvg & \rmfamily 0.130117  & \rmfamily 0.129255 \\
  \rmfamily CoCitationSampling+CitMod & \rmfamily \textbf{0.145235}  & \rmfamily \textbf{0.142829} \\
  \hline
\end{tabular}
\label{tab:high_hidden}
\end{table}

Intuitively this behavior makes sense. With a low hidden ratio, the
query passed to the system very clearly specifies the need of the user
and a method like PaperRank can directly take advantage of that. With
a high hidden ratio, the query of the user is vague and then the
embedding more accurately capture the structure of that region of the literature.

\section{Conclusion and Future Work}
\label{sec:ccl}
In this paper, we presented the node embedding on graph for the citation recommendation task.
Besides the random walk stream based sampling strategy which encodes the general graph structural
information, we proposed a task specific sampling strategy using co-citation relationships.

In order to evaluate the embedding results on citation recommendation task, we need a scheme to
score the candidate papers based on a set of seed papers. Therefore, we designed three embedding
based rankings: $SimAvg$,$SimWgd$ and $SimRef$ and one model based ranking: $CitMod$.

The experimental results show the co-citation sampling strategy outperforms the random walks based
sampling strategy on all ranking schemes, and the model based ranking outperforms embedding based
rankings for both sampling schemes.


Then we demonstrated that graph embedding is a robust approach for citation recommendation when hidden ratio changes, while
the performance of PR and CF drops a lot when hidden ratio is becoming large. In general, when the hidden ratio is
small, classic methods are better, but when the size of seed paper set is small, co-citation sampling based embedding
is a better choice.
CCS reaches a better performance than both PR and CF when the hidden ratio is large.
This is directly applicable to recommendation services that can use a different algorithm depending on the size of the query.

In the future, we would like to explore how graph visualization helps citation recommendation systems.
Graph-based organization shows
some advantages compared with list-based organization. Users rarely browse the papers appearing after the first page
for list-based organization.  But a graph visualization can show more papers at once and can show the citation structures which should enable
users easily to find the papers they are interested in. We expect that graph embedding techniques will help to highlight the structure of recommended papers and help the user quickly make sense of the different aspect of the literature.

\begin{acknowledgements}
This material is based upon work supported by the National Science Foundation under Grant No. 1652442.
\end{acknowledgements}


\bibliographystyle{spmpsci}
\bibliography{template}

\end{document}